\documentclass{emulateapj}

\def\tobs{t_{\rm obs}}

\def\nuobs{\nu_{\rm obs}}

\def\nuen{\nu_{\rm en}}

\def\be{\begin{equation}}
\def\ee{\end{equation}}
\def\beq{\begin{eqnarray}}
\def\eeq{\end{eqnarray}}

\begin{document}

\title{On the Curvature Effect of a Relativistic Spherical Shell}

\author{Z. Lucas Uhm\altaffilmark{1,3}, Bing Zhang\altaffilmark{1,2,3}}

\altaffiltext{1}{Kavli Institute for Astronomy and Astrophysics, Peking University, Beijing 100871, China}
\altaffiltext{2}{Department of Astronomy, School of Physics, Peking University, Beijing 100871, China}
\altaffiltext{3}{Department of Physics and Astronomy, University of Nevada, Las Vegas, NV 89154, USA}

\begin{abstract}
We consider a relativistic spherical shell and calculate its spectral flux as 
received by a distant observer. Using two different methods, we derive a simple 
analytical expression of the observed spectral flux and show that the well-known 
relation $\hat \alpha = 2+\hat \beta$ (between temporal index $\hat \alpha$ and 
spectral index $\hat \beta$) of the high-latitude emission is achieved naturally 
in our derivation but holds only when the shell moves with a constant Lorentz 
factor $\Gamma$. 
Presenting numerical models where the shell is under acceleration or deceleration, 
we show that the simple $\hat \alpha = 2+\hat \beta$ relation is indeed deviated 
as long as $\Gamma$ is not constant. For the models under acceleration, 
we find that the light curves produced purely by the high-latitude emission decay 
initially much steeper than the constant $\Gamma$ case and gradually resume 
the $\hat \alpha = 2+\hat \beta$ relation in about one and half orders of magnitude 
in observer time. For the models under deceleration, the trend is opposite. 
The light curves made purely by the high-latitude emission decay initially shallower 
than the constant $\Gamma$ case and gradually resume the relation 
$\hat \alpha = 2+\hat \beta$ in a similar order of magnitude in observer time. 
We also show that how fast the Lorentz factor $\Gamma$ of the shell increases or 
decreases is the main ingredient determining the initial steepness or shallowness 
of the light curves. 
\end{abstract}


%
%

\section{Introduction} \label{section:introduction}

In the astrophysical phenomena involving relativistic jets, such as gamma-ray bursts (GRBs), 
the relativistic beaming of radiation plays an important role and leads to an interesting 
effect especially when combined with a non-planar geometry. For a jet with spherical geometry, 
the emission from a jet location that has higher latitude than the line of sight takes longer 
time to reach an observer than the emission along the line of sight. 
Thus, although emitted simultaneously from the jet, this so-called ``high-latitude 
emission'' spreads out along the time axis as received by the observer. Also, due to the 
relativistic beaming effect, the emission from higher latitudes has progressively smaller 
Doppler factor, so that the observed flux density decays rapidly with the observer time. 
These two aspects of the high-latitude emission are known as the ``curvature effect'' of 
a relativistic spherical shell.

If the photon spectrum has a power-law shape in the fluid frame co-moving with the spherical shell, 
the high-latitude emission from the shell produces an observed spectral flux 
$F_{\nuobs}^{\, \rm obs}$ at an observed frequency $\nuobs$, such that it satisfies a simple 
relation between the temporal index $\hat \alpha$ and the spectral index $\hat \beta$, 
\begin{equation}
\label{eq:alpha_beta}
\hat \alpha = 2+\hat \beta,
\end{equation}
in the convention of 
$F_{\nuobs}^{\, \rm obs} \propto \tobs^{-\hat \alpha}\, \nuobs^{-\hat \beta}$, 
where $\tobs$ is the observer time. 
This relation was first correctly derived by \cite{kumar00}, and later verified by
several authors both analytically \citep[e.g.,][]{dermer04} and numerically 
\citep[e.g.,][]{dyks06}\footnote{The same expression (\ref{eq:alpha_beta}) was also
presented earlier by \cite{fenimore96}. However, their spectral index was defined
for the photon number index rather than the flux density index. As a result, 
their relation gives $\hat \alpha=3+\hat \beta$ in our notation, 
which is off by 1. See also \cite{dermer04}.}.

In reality, when the spherical shell emits continuously, the observed spectral flux 
$F_{\nuobs}^{\, \rm obs}$ does not follow Equation (\ref{eq:alpha_beta}) 
since the emission from higher latitudes has smaller Doppler boosting and 
is buried under the continuous emission from the jet emitted at later times. 
In this case, the temporal evolution of $F_{\nuobs}^{\, \rm obs}$ is mainly determined 
by the time evolution of the jet power along the observer's line of sight. However, 
for the jets with rapid variability like in the GRB jets, one may consider a situation 
where the emission from the jet ceases abruptly. In such a case, 
the observed spectral flux can be purely produced by the high-latitude emission, 
and thus, the curvature effect of the spherical shell shapes the observed light curves. 
This effect has been invoked to interpret the steep decay phase of early X-ray
afterglow of GRBs \citep{zhang06,zhangbb09,genet09} and the decay segment
of the X-ray flares following GRBs \citep{liang06}.

In this paper, we present a simple analytical derivation of the observed spectral flux 
$F_{\nuobs}^{\, \rm obs}$ from a relativistic spherical shell, for the case of an arbitrary 
shape of photon spectrum in the fluid frame. 
We derive the same expression while employing two different approaches: 
(1) the {\it emitted} spectral power of the electrons in the shell and (2) the {\it received} spectral 
power of the electrons and an integration over the equal-arrival time surface. 
Then, we show 
that Equation (\ref{eq:alpha_beta}) for the high-latitude emission 
is naturally satisfied in our derivation but holds only in the case where the shell expands 
with a constant value of the bulk Lorentz factor. Presenting numerical models 
where the shell is under acceleration or deceleration, we show that the high-latitude 
emission indeed deviates from Equation (\ref{eq:alpha_beta}) for an accelerating or 
decelerating shell. We also discuss its possible implications in the context of GRB 
observations. 
During the afterglow phase, the emission region is known to be under 
deceleration \citep{meszarosrees97,sari98}. 
On the other hand, during the prompt emission phase, the emission region may be 
during an acceleration phase, 
if the prompt emission is powered by 
dissipating magnetic energy in a Poynting flux dominated jet \citep[e.g.,][]{zhangyan11}.

%
%

\section{Curvature effect of a spherical shell}

We first analytically derive the curvature effect of a spherical shell using two
different methods.

Consider a thin spherical shell of radius $r$ at time $t$ expanding with a bulk Lorentz 
factor $\Gamma$ in the lab frame, which was initially ejected at time $t=0$ from a 
central explosion at radius $r=0$. An observer located at a large cosmological distance 
from the shell sets the observer time $\tobs$ equal to zero upon receiving the very first 
photon emitted from the center at time $t=0$. Then, a photon emitted from the shell at 
time $t$ from a location of a polar angle $\theta$ with respect to the observer's line of 
sight will be detected by the observer at observer time 
\be
\label{eq:tobs}
\tobs = \left(t-\frac{r}{c}\, \mu \right) (1+z),
\ee
where $\mu \equiv \cos \theta$, $c$ is the speed of light, 
and $z$ is the redshift of the explosion.

Also, consider a total of $N$ electrons uniformly distributed in the shell and 
assume that, in the fluid frame co-moving with the shell, an electron of a 
Lorentz factor $\gamma_e$ has its spectral power 
$P_{\nu^{\prime}}^{\prime} \equiv dP^{\prime}/d\nu^{\prime}$ 
at frequency $\nu^{\prime}$ as\footnote{Our general approach described here applies to
an unspecified non-thermal radiation mechanism. The curvature effect does not depend on
the explicit radiation mechanism. We will however introduce the synchrotron radiation in
Section~\ref{section:examples} when presenting our numerical calculations.} 
\be
\label{eq:spec_power_single}
P_{\nu^{\prime}}^{\prime} (\nu^{\prime}) \propto H(x)
\quad
\mbox{with}
\quad
x=\nu^{\prime}/\nu_0^{\prime},
\ee
so that the photon spectrum of the electron is shaped by the functional form of $H(x)$ 
and is located at a characteristic frequency $\nu_0^{\prime}$. 
At every location in the shell, the electrons form a spectrum $dN_e/d\gamma_e$ 
in their energy space. Thus, the photon spectrum emitted from those electrons needs 
to be evaluated as a convolution of Equation (\ref{eq:spec_power_single}) with 
the electron spectrum $dN_e/d\gamma_e$, which would then yield a spectral shape different 
from that of $H(x)$. 
In order to describe this photon spectrum arising from a group or ``ensemble'' of 
electrons placed together, we introduce here a functional form $H_{\rm en} (x)$ with 
$x=\nu^{\prime}/\nuen^{\prime}$, which is then located at an ensemble frequency 
$\nuen^{\prime}$. For instance, the function $H_{\rm en} (x)$ may have a GRB ``Band-function'' 
shape \citep{band93} or simply a power-law shape. 
Dividing the spectral luminosity emitted from all the electrons within the ensemble 
by the number of electrons included there, we now assume that the spectral power 
of each electron within the ensemble may also be on average described by the same functional 
form. In other words, we propose 
\be
\label{eq:spec_power_single_ensemble}
P_{\nu^{\prime}}^{\prime} (\nu^{\prime}) = 
P_0^{\prime}\, H_{\rm en}(x)
\quad
\mbox{with}
\quad
x=\nu^{\prime}/\nuen^{\prime},
\ee
where $P_0^{\prime}$ is a measure of the spectral power of a single electron 
in the fluid frame. Note that $P_0^{\prime}$ and $\nuen^{\prime}$ here are not necessarily 
constant and can evolve in time. 
With the set-up depicted in Equation (\ref{eq:spec_power_single_ensemble}), we can efficiently 
investigate solely the relativistic curvature effect of a spherical shell as we proceed below, 
without invoking a detailed shape of the electron spectrum or a specific radiation process.

Consider again an electron located in the shell with a polar angle $\theta$ with respect to 
the observer's line of sight. Then, a photon emitted from the electron in the direction of 
the observer has the same angle $\theta$ with the radial bulk motion of the electron, and 
thus appears with a frequency $\nu$ in the lab frame,
\be
\nu (\theta) = \nu^{\prime}\, [\Gamma (1-\beta \mu)]^{-1}, 
\ee
while the same photon has a frequency $\nu^{\prime}$ in the fluid frame. 
Here, $\beta$ is given by $\beta=(1-\Gamma^{-2})^{1/2}$. 
Provided that the electron emits photons isotropically\footnote{
A possible anisotropic emission in the fluid frame was considered in \cite{beloborodov11}. 
The authors showed that the spreading effect in the light curves due to the high-latitude emission 
can be reduced if emission is anisotropic in the fluid frame. This effect may appear similar to 
what we show in the current paper in the case of an accelerating shell (see Section~\ref{section:examples}).} 
in the fluid frame, 
the spectral energy $\delta E_{\nu}$ {\it emitted} by the electron at 
frequency $\nu$ during a time interval $\delta t$ into a solid angle $\delta \Omega$ 
in the direction of the observer is given in the lab frame as 
\be
\delta E_{\nu} =
\frac{\delta t\, \delta \Omega}{\Gamma^3 (1-\beta \mu)^2}\, 
\frac{1}{4\pi}\, P_0^{\prime}\, H_{\rm en}(\nu/\nuen),
\ee
where $\nuen \equiv \nuen^{\prime}\, [\Gamma (1-\beta \mu)]^{-1}$.

Now consider a thin ring in the shell in a polar angle range between $\theta$ 
and $\theta + \delta \theta$. Since the number of electrons contained in the ring 
is given by $(|\delta \mu|/2) N$, 
the spectral energy $\delta \tilde E_{\nu}$ emitted from the ring at 
frequency $\nu$ during $\delta t$ into $\delta \Omega$ in the 
direction of the observer reads in the lab frame as
\be
\delta \tilde E_{\nu} = 
\delta E_{\nu}\, (|\delta \mu|/2) N.
\ee
Here, the tilde indicates the ring.
In reality, the electrons in the ring emit photons continuously as the shell expands. 
However, as described in \cite{uhm12}, we may view the emission from the shell as 
a series of ``flashes''. We assume that the electrons in the ring accumulate their 
emission between any two consecutive flashes (separated by a time interval $\delta t$) 
and emit all the accumulated energy instantaneously like a flash at the end of each 
time interval.

When the spectral energy $\delta \tilde E_{\nu}$ of the ring is 
released into $\delta \Omega$ as a flash, the ring's thickness 
(between $\theta$ and $\theta+\delta \theta$) introduces a time interval 
$\delta \tilde t = (r/c)\, |\delta \mu|$ in the lab frame along the observer's 
line of sight \citep{uhm12}. Hence, the spectral luminosity of the ring at frequency 
$\nu$, which is shone into $\delta \Omega$ in the direction of the observer, 
reads in the lab frame as 
\beq
\delta L_{\nu} 
&=& \frac{\delta \tilde E_{\nu}}{\delta \tilde t} 
 =  \frac{c}{2r}\, N\, \delta E_{\nu} \nonumber \\
\label{eq:dLlab}
&=& \frac{1}{4\pi}\, \frac{c}{2r}\, 
\frac{N\, \delta t\, \delta \Omega}{\Gamma^3 (1-\beta \mu)^2}\, 
P_0^{\prime}\, H_{\rm en}(\nu/\nuen),
\eeq
where the ring's thickness, i.e., $\delta \mu$ cancels out, and 
thus we drop out a tilde from $\delta L_{\nu}$.

The photons emitted into $\delta \Omega$ at frequency $\nu$ are redshifted 
while traveling and would be seen by the observer at an observed frequency 
\be
\nuobs=\nu/(1+z).
\ee
The observed spectral flux at frequency $\nuobs$ is then given as 
\beq
\delta F_{\nuobs}^{\, \rm obs}
&=& \frac{(1+z)\,\delta L_{\nu}}{D_{\rm L}^2\delta \Omega} \nonumber \\
\label{eq:observed_flux}
&=& \frac{1+z}{4\pi D_L^2} \frac{c}{2r} 
\frac{N P_0^{\prime}\, \delta t}{\Gamma^3 (1-\beta \mu)^2} 
H_{\rm en}(\nuobs/\nuen^{\rm obs}),
\eeq
where $D_L$ is the luminosity distance to the shell from the observer, and 
$\nuen^{\rm obs} \equiv \nuen/(1+z)$. 
Note that the solid angle $\delta \Omega$ also cancels out here. 
Finding $\mu$ from Equation (\ref{eq:tobs}), 
\be
\label{eq:mu}
\mu = \frac{c}{r} \left(t - \frac{\tobs}{1+z} \right),
\ee
we now have an integral for $F_{\nuobs}^{\, \rm obs}$ in terms of $\tobs$ and $\nuobs$, 
\be
\label{eq:observed_flux_integral}
F_{\nuobs}^{\, \rm obs}
= \frac{1+z}{4\pi D_L^2} \int \frac{c}{2r} 
\frac{N P_0^{\prime} H_{\rm en}( (1+z) \Gamma (1-\beta \mu) \nuobs/\nuen^{\prime})}
{\Gamma^3 (1-\beta \mu)^2} dt.
\ee
Here, we have used $\nuen^{\rm obs} = \nuen^{\prime}\, [(1+z)\, \Gamma (1-\beta \mu)]^{-1}$. 
Note that the redshift factor $z$ enters Equation (\ref{eq:observed_flux_integral}) only 
through the combinations $\tobs/(1+z)$ and $\nuobs (1+z)$ as well as the overall normalization. 
Therefore, redshift only plays a global role in shaping 
the observed spectral flux $F_{\nuobs}^{\, \rm obs}$ 
in a 3-D space ($\tobs$, $\nuobs$, $F_{\nuobs}^{\, \rm obs}$).

It is widely known that the high-latitude emission from a spherical shell satisfies 
Equation (\ref{eq:alpha_beta}) for the convention 
$F_{\nuobs}^{\, \rm obs} \propto \tobs^{-\hat \alpha}\, \nuobs^{-\hat \beta}$ \citep{kumar00}. 
We point out that this relation $\hat \alpha=2+\hat \beta$ is naturally achieved in our derivation above. 
For $H_{\rm en} (x) \propto x^{-\hat \beta}$, Equation (\ref{eq:observed_flux_integral}) gives 
\be
\label{eq:high_latitude_emission}
F_{\nuobs}^{\, \rm obs} \propto 
\int \frac{N P_0^{\prime} \nuen^{\prime\, \hat \beta} \nuobs^{-\hat \beta}}
{r \Gamma^{3+\hat \beta} (1-\beta \mu)^{2+\hat \beta}}\, dt.
\ee
Consider now a constant value of $\Gamma$, which ensures that $r=c\beta t$ is satisfied. 
Equation (\ref{eq:tobs}) then becomes 
\be
\label{eq:tobs_constant_gamma}
\tobs = t\, (1-\beta \mu) (1+z). 
\ee
Combining Equations (\ref{eq:high_latitude_emission}) and (\ref{eq:tobs_constant_gamma}), 
we have
\be
\label{eq:high_latitude_constant_gamma}
F_{\nuobs}^{\, \rm obs} \propto 
\tobs^{-(2+\hat \beta)}\, \nuobs^{-\hat \beta}
\int N P_0^{\prime} \nuen^{\prime\, \hat \beta}\, 
t^{1+\hat \beta}\, dt.
\ee
While the shell is still emitting, 
the integral in Equation (\ref{eq:high_latitude_constant_gamma}) varies in time, but 
once the emission from the shell is turned off, the integral becomes a constant value. 
Therefore, the observed spectral flux $F_{\nuobs}^{\, \rm obs}$ beyond the turn-off point, 
which is then produced purely by the high-latitude emission, 
satisfies the relation $\hat \alpha=2+\hat \beta$. 
We stress, however, that this relation $\hat \alpha=2+\hat \beta$ does not hold any longer 
when $\Gamma$ evolves in time because Equation (\ref{eq:tobs_constant_gamma}) becomes invalid; 
see also Section~\ref{section:examples}.

We now present an alternative derivation of Equation (\ref{eq:observed_flux_integral}), 
by making use of an equal-arrival time surface (EATS). 
Consider an electron located in the shell at time $t$ with a polar angle $\theta$ 
with respect to the observer's line of sight. 
Its {\it received} spectral power at frequency $\nu$ into a solid angle 
$\delta \Omega$ in the direction of the observer is given in the lab frame as 
\be
\label{eq:dPlab_rec}
\delta P_{\nu}^{\, \rm rec} =
\frac{\delta \Omega}{\Gamma^3 (1-\beta \mu)^3}\, 
\frac{1}{4\pi}\, P_0^{\prime}\, H_{\rm en}(\nu/\nuen).
\ee
The photons emitted from this electron at time $t$ will be received by the observer 
at an observer time $\tobs$ (given by Equation (\ref{eq:tobs})). 
Now we integrate over the EATS of this $\tobs$ between $t$ and $t+\delta t$, 
by counting the number of electrons that contribute to the same $\tobs$. 
During the time interval $\delta t$, the shell travels a distance of 
$c\beta\, \delta t$ and has a radius of $r+c\beta\, \delta t$ at time $t+\delta t$. 
The definition of EATS of this $\tobs$ reads
\be
\label{eq:EATS}
r \cos \theta + c\, \delta t = 
(r+c\beta\, \delta t) \cos (\theta-\delta \theta),
\ee
where $\theta-\delta \theta$ is the polar angle of EATS of this $\tobs$ 
at time $t+\delta t$. 
Note that during the shell's expansion for $\delta t$, the polar angle 
of EATS of this $\tobs$ decreases by an amount of $\delta \theta$. 
Since $\cos (\theta-\delta \theta) \simeq \cos \theta + \sin \theta\, \delta \theta$, 
Equation (\ref{eq:EATS}) gives 
\be
\sin \theta\, \delta \theta \simeq 
\frac{c}{r}\, (1-\beta \cos \theta)\, \delta t.
\ee
The number of electrons contained on the EATS of this $\tobs$ between $r$ and 
$r+c\beta\, \delta t$ (or equivalently between $\theta$ and $\theta-\delta \theta$) 
is equal to the number of electrons contained in the shell in the polar angle range 
between $\theta-\delta \theta$ and $\theta$, which is given by 
\be
(|\delta \mu|/2) N = 
\frac{1}{2} (\sin \theta\, \delta \theta) N = 
\frac{c}{2r} (1-\beta \mu) N\, \delta t. 
\ee
These electrons, contained on the EATS of this $\tobs$ in the time range 
between $t$ and $t+\delta t$, contribute to the same $\tobs$ and gives the spectral 
luminosity $\delta L_{\nu}$ at frequency $\nu$ as follows 
\be
\label{eq:dLlab_EATS}
\delta L_{\nu} = 
\left[ \frac{c}{2r}\, (1-\beta \mu) N\, \delta t \right] 
\delta P_{\nu}^{\, \rm rec}.
\ee
Note that Equation (\ref{eq:dLlab_EATS}) becomes identical to the result above, 
Equation (\ref{eq:dLlab}), when Equation (\ref{eq:dPlab_rec}) is substituted in. 
Hence, we arrive at Equation (\ref{eq:observed_flux_integral}) again.

%
%

\section{Numerical examples} \label{section:examples}

We consider a spherical shell at redshift $z=1$. For its luminosity distance $D_L$ 
from the observer, we adopt a flat $\Lambda$CDM universe with the parameters 
$H_0=71$ km $\mbox{s}^{-1}$ $\mbox{Mpc}^{-1}$, 
$\Omega_{\rm m}=0.27$, and $\Omega_{\Lambda}=0.73$ (the concordance model). 
The number of electrons in the shell $N$ is assumed to increase at a constant 
injection rate $R_{\rm inj} \equiv dN/dt^{\prime} = 10^{45} ~ \mbox{s}^{-1}$ 
from an initial value $N = 0$, 
where $t^{\prime}$ is the time measured in the co-moving fluid frame. 
For the functional form of $H_{\rm en} (x)$, we take a simple power-law shape 
$H_{\rm en} (x) = x^{-\hat \beta}$ with a spectral index $\hat \beta = 1$. 
Regarding the choice of $P_0^{\prime}$ and $\nuen^{\prime}$, 
having the synchrotron radiation in mind, 
we adopt the followings from the synchrotron theory \citep{rybicki79}\footnote{
Assuming that the electrons have an isotropic distribution of their pitch-angle $\alpha$ 
in the fluid frame, we take an average over the distribution so that 
$<\sin \alpha> \,\, = (4\pi)^{-1} \int \sin \alpha\, d\Omega_{\alpha} =
(1/2) \int_0^{\pi} \sin^2 \alpha\, d\alpha = \pi/4$.} 
\be
P_0^{\prime} = \frac{3 \sqrt{3}}{32}\, \frac{m_e c^2\, \sigma_T B}{q_e},
\quad
\nuen^{\prime}=\frac{3}{16}\, \frac{q_e B}{m_e c}\, \gamma_{\rm inj}^2.
\ee
Here, $m_e$ and $q_e$ are the mass and charge of the electron, respectively, 
and $\sigma_T$ is the Thomson cross section. 
The magnetic field strength $B$ in the shell and the injection Lorentz factor 
$\gamma_{\rm inj}$ of the electrons are measured in the fluid frame. 
Choosing $B=30$ G and $\gamma_{\rm inj}=5 \times 10^4$, we place the ensemble 
spectrum at around $h \nuen^{\prime} \simeq 1$ keV in the fluid frame.
Such a set of parameters are the right ones to reproduce the observed prompt 
emission spectra of GRBs \citep{uhmzhang14,zhangbb15}.

We present nine numerical models, for which everything given above 
remains the same. The first model we present (named [1a]) is under constant bulk 
motion with $\Gamma=300$, while the other eight models are under acceleration or 
deceleration with $\Gamma$ in a power-law form in radius: 
$\Gamma(r)=\Gamma_0\, (r/r_0)^s$ with $r_0=10^{14}$ cm. 
The second model (named [2a]) is under acceleration with $\Gamma_0=10^2$ and $s=0.4$, and 
the third model (named [3a]) is under deceleration with $\Gamma_0=10^3$ and $s=-0.4$. 
We begin our calculations at radius $r_0$ (and at time $t_0=r_0/(c\beta)$ for $\Gamma=300$) 
and turn off the emission of the shell 
at $\hat \tobs=3$ s. Here, $\hat \tobs$ is defined by 
$\hat \tobs = (1+z) \int dt/(2\Gamma^2)$, and measures the observed time of 
photons emitted with $\theta=0$ along the observer's axis\footnote{
Since $1-\beta \simeq 1/(2\Gamma^2)$, we have 
$\hat \tobs \simeq (1+z) \int (1-\beta) dt = (1+z) (t-r/c)$, 
which is the same as the observer time $\tobs$ (in Equation (\ref{eq:tobs})) 
for $\theta=0$.}. 
Note that the same turn-off time ($\hat \tobs=3$ s) corresponds to a different 
turn-off radius for each of these three models [1a], [2a], and [3a] since 
they have different $\Gamma(r)$ profiles.

\begin{figure}
\begin{center}
\includegraphics[width=17cm]{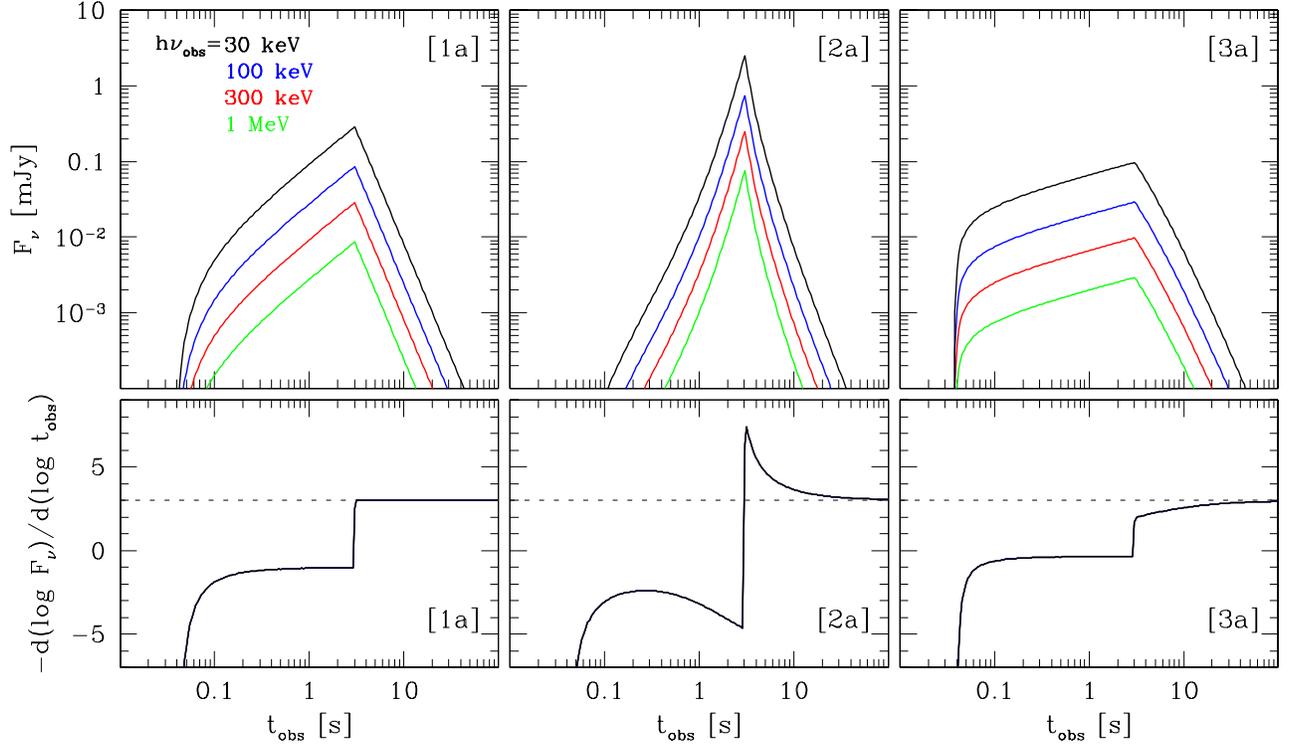}
\caption{
Light curves for models [1a], [2a], and [3a]. Top panels show the 
model light curves at 30 keV (black), 100 keV (blue), 300 keV (red), 
and 1 MeV (green), respectively, while the bottom panels show the temporal 
index $\hat \alpha$ of these four light curves. We turn off the emission 
of the spherical shell at $\hat \tobs=3$ s, so that the light curves 
beyond this turn-off time display the high-latitude emission from the shell. 
The dotted line in the bottom panels represents the relation 
$\hat \alpha = 2+\hat \beta$ for the spectral index $\hat \beta=1$. 
The model [1a] is under constant bulk motion, the model [2a] is under 
acceleration, and the model [3a] is under deceleration. 
}
\label{fig:1a2a3a_fnu_lc}
\end{center}
\end{figure}

Figure~\ref{fig:1a2a3a_fnu_lc} shows the resulting light curves of 
models [1a], [2a], and [3a]. In the upper panels, we show the observed spectral 
flux $F_{\nuobs}^{\, \rm obs}$ as a function of observer time $\tobs$ 
at $h\nuobs=30$ keV (black), 100 keV (blue), 300 keV (red), and 1 MeV (green), 
respectively, and in the lower panels, we show the temporal index 
$\hat \alpha = -d(\log F_{\nuobs}^{\, \rm obs})/d(\log \tobs)$ of these four 
light curves. The dotted line in the lower panels represents the relation 
$\hat \alpha = 2+\hat \beta$ for the spectral index $\hat \beta=1$. 
The light curves in all three models rise initially 
(since $N$ increases with time), peak at the turn-off time at 3 s, and 
then decay subsequently beyond that time, displaying a high-latitude emission 
of the shell. For the model [1a] with a constant value of $\Gamma$, 
it is noted that the $\hat \alpha$ curve agrees with the expected relation 
$\hat \alpha = 2+\hat \beta$ beyond the turn-off time. However, for the model [2a] 
under acceleration, the $\hat \alpha$ curve beyond the turn-off time indicates that 
the light curves produced purely by the high-latitude emission decay initially 
much steeper than in the model [1a] and then gradually resume the relation 
$\hat \alpha = 2+\hat \beta$ in about one and half orders of magnitude 
in observer time. For the model [3a] under deceleration, the trend is in the 
opposite direction. The light curves produced purely by the high-latitude emission 
beyond the turn-off time are initially shallower than in the model [1a]
and gradually resume the relation $\hat \alpha = 2+\hat \beta$ 
in about one and half orders of magnitude in observer time.

\begin{figure}
\begin{center}
\includegraphics[width=11cm]{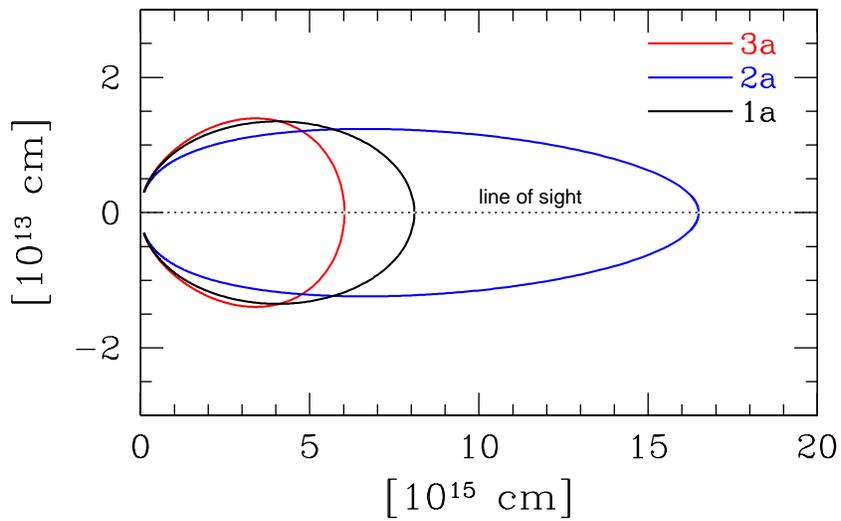}
\caption{
Equal-arrival time surface (EATS) of the models [1a], [2a], and [3a]. 
These EATS's correspond to the observer time $\tobs=3$ s.
}
\label{fig:1a2a3a_eas}
\end{center}
\end{figure}

We also calculate the EATS (of contributing to $\tobs=3$ s) for these three 
models [1a], [2a], and [3a] and show them in Figure~\ref{fig:1a2a3a_eas}. 
As compared to the ellipsoidal shape of EATS of the model [1a], 
the EATS of the model [2a] (under acceleration) is elongated 
along the line of sight further on the side of larger radii. 
On the other hand, the EATS of the model [3a] (under deceleration) 
is elongated lesser on the side of larger radii, as also shown 
in previous publications \citep[e.g.,][]{sari98b}.
This difference in the shape of three EATS's 
can help visualize our finding in Figure~\ref{fig:1a2a3a_fnu_lc}.

\begin{figure}
\begin{center}
\includegraphics[width=11cm]{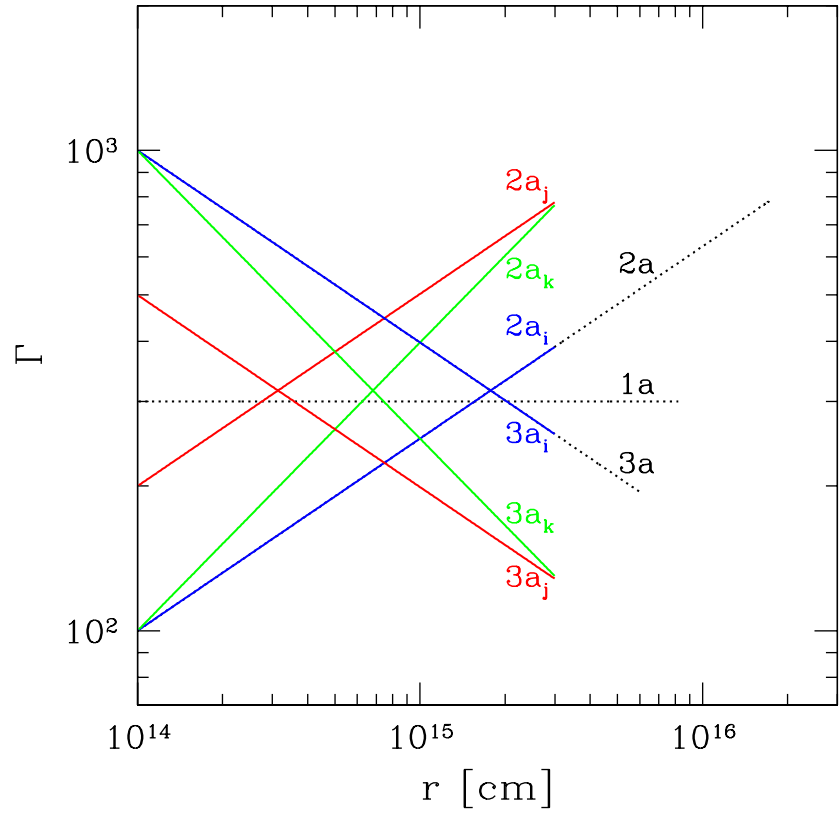}
\caption{
Lorentz factor $\Gamma$ of the shell shown as a function of radius $r$
for the nine numerical models presented. 
}
\label{fig:gb_all}
\end{center}
\end{figure}

In order to better understand this deviation from the expected relation 
$\hat \alpha = 2+\hat \beta$, we now make the following three variations 
on the model [2a] and another three variations on the model [3a]. 
Firstly, we would like to see if the turn-off radius matters. 
Thus, for the model [2a$_i$], we take the same profile of $\Gamma(r)$ as 
in the model [2a] but turn off the emission of the shell at a smaller radius 
$r_{\rm off} = 3\times 10^{15}~\mbox{cm}$ than in the model [2a]. 
For the model [2a$_j$], we keep everything the same as in the model [2a$_i$] 
but increase $\Gamma_0$ by a factor of 2. 
For the model [2a$_k$], everything is the same as in the model [2a$_i$] 
but $s$ is changed to a higher value $s=0.6$. 
Three variations on the model [3a] are made in the same way. 
The model [3a$_i$] has the same profile of $\Gamma(r)$ as in the model [3a] 
but has the turn-off radius $r_{\rm off}$. 
The model [3a$_j$] has a smaller $\Gamma_0$ by a factor of 2 when compared 
to the model [3a$_i$]. The model [3a$_k$] has a lower value of $s=-0.6$ 
as compared to the model [3a$_i$]. 
In Figure~\ref{fig:gb_all}, we show all of these six variations, 
together with the previous models [1a], [2a], and [3a].\footnote{ 
Here, we note again that the turn-off radius of the models [1a], [2a], and [3a] 
was determined individually by setting the turn-off time $\hat \tobs=3$ s. 
}

\begin{figure}
\begin{center}
\includegraphics[width=17cm]{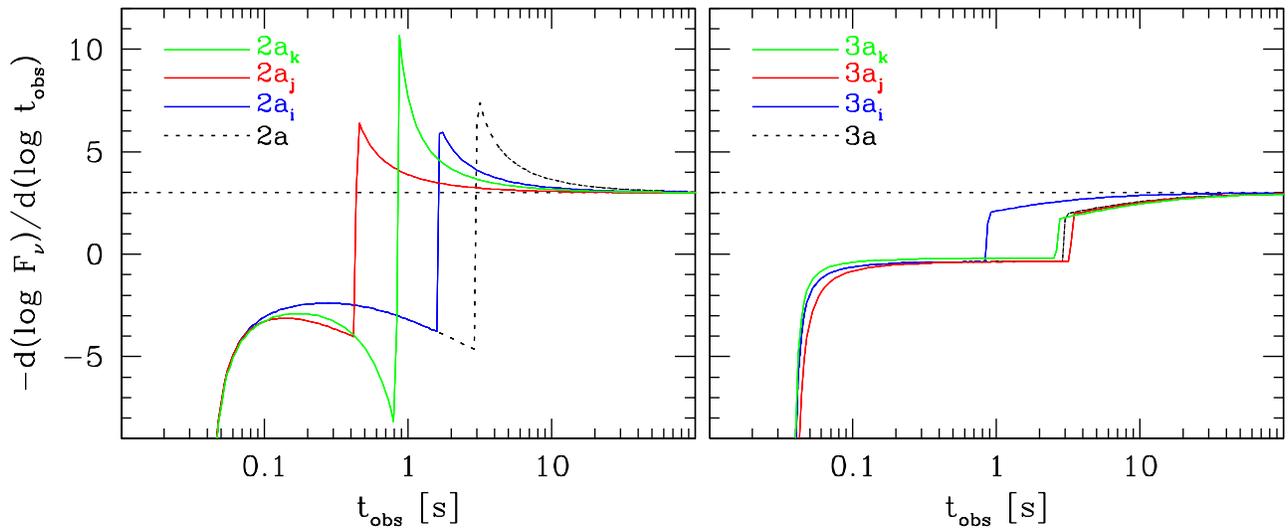}
\caption{
Temporal index $\hat \alpha$ shown as a function of observer time $\tobs$ 
for the four numerical models under acceleration (Left) and for the four 
models under deceleration (Right). 
}
\label{fig:alpha}
\end{center}
\end{figure}

We repeat our calculations for these new models and show the $\hat \alpha$ curve 
of each model in Figure~\ref{fig:alpha}. 
From the left panel, we conclude that, in the case of an accelerating spherical shell, 
the steepness of the light curves beyond the turn-off point depends weakly on 
the turn-off radius and the value $\Gamma_0$, but responds most sensitively 
to the value of the acceleration index $s$. The higher the value $s$, the steeper the light curves. 
In the case of a decelerating spherical shell, the right panel shows that 
the shallowness of the light curves beyond the turn-off point is nearly insensitive 
to the turn-off radius and the value $\Gamma_0$, but depends weakly on 
the value of the deceleration index $s$. The lower the value $s$, the shallower the light curves.

%
%

\section{Conclusions and Discussion}

In this paper, we consider a relativistic spherical shell expanding with a bulk 
Lorentz factor $\Gamma$ and calculate the spectral flux received by a distant 
observer located at a large cosmological distance. Assuming an arbitrary shape of 
photon spectrum in the fluid frame co-moving with the shell, we present a simple 
analytical derivation of the observed spectral flux $F_{\nuobs}^{\, \rm obs}$ 
in terms of observer time $\tobs$ and observed frequency $\nuobs$. 
In particular, we derive the same expression while making use of two different 
approaches: (1) the emitted spectral power of the electrons and (2) the received 
spectral power of the electrons and an integration over the equal-arrival time 
surface. It is known that the high-latitude emission from a spherical shell 
satisfies a relation $\hat \alpha = 2+\hat \beta$ between the temporal 
index $\hat \alpha$ and the spectral index $\hat \beta$. We show that this 
relation is naturally achieved in our derivation but holds only in the case of 
a constant value of $\Gamma$.

We present nine numerical models: One model under constant bulk motion (named [1a]), 
four models under acceleration (named [2a], [2a$_i$], [2a$_j$], and [2a$_k$]), 
and another four models under deceleration (named [3a], [3a$_i$], [3a$_j$], and [3a$_k$]). 
Calculating the light curves at four different energy bands and finding 
the temporal index $\hat \alpha$ of those light curves for each model, 
we show that the relation $\hat \alpha = 2+\hat \beta$ is indeed satisfied 
only for the first model [1a]. For the models under acceleration, we find that 
the light curves produced purely by the high-latitude emission decay initially 
much steeper than in the model [1a] and gradually resume the relation 
$\hat \alpha = 2+\hat \beta$ in about one and half orders of magnitude in observer time. 
For the models under deceleration, the trend is opposite. 
We show that, in the case of a decelerating spherical shell, 
the light curves produced purely by the high-latitude emission decay initially 
shallower than in the model [1a] and gradually resume the relation 
$\hat \alpha = 2+\hat \beta$ again in about one and half orders of magnitude in observer time. 
More specifically, we find that, for a shell under acceleration, the initial 
steepness of the high-latitude emission depends most sensitively on how fast 
the Lorentz factor $\Gamma$ increases, but also depends weakly 
on the value $\Gamma$ itself and the radius where we turn off the emission of the shell. 
In the case of a decelerating shell, we show that the initial shallowness of 
the high-latitude emission depends weakly on how fast the Lorentz factor 
$\Gamma$ decreases, but is nearly insensitive to the value $\Gamma$ itself 
and the radius where the emission of the shell is turned off.

This departure from the relation $\hat \alpha = 2+\hat \beta$ may find applications to 
many aspects of GRB observations. 
It is well known that during the afterglow phase, the emission region is under 
deceleration \citep{meszarosrees97,sari98}. 
If the afterglow emission from the blast wave ceases abruptly, e.g., when the blast wave 
enters a density void as originally envisaged by \cite{kumar00}, 
the observed light curves would be shaped by the high-latitude emission 
arising from the blast that has been decelerating. 
More interestingly, during the prompt emission phase, the emission region may be 
during an acceleration phase, 
if the prompt emission is powered by dissipating magnetic energy via internal
collision-induced magnetic reconnection and turbulence (ICMART, \citealt{zhangyan11}).
This is because ICMART events are expected to happen when the bulk magnetization 
parameter $\sigma$ (ratio between Poynting flux and matter flux) is above unity,
so that the outflow is still during an acceleration phase 
\citep[e.g.,][]{komissarov09,granot11}. 
During the ICMART process, $\sigma$ is expected to drop rapidly. 
Part of the dissipated magnetic energy would be converted to the kinetic energy of the outflow, 
giving rise to an extra acceleration to the outflow \citep[e.g.,][]{zhangzhang14}. 
Identifying a deceleration signature in the afterglow emission can directly confirm 
the deceleration nature of the afterglow. Also, identifying an acceleration feature 
in the prompt emission would have profound implications for our understanding of 
the jet composition and energy dissipation mechanism of the prompt emission. 
An application of the theory presented here to GRB data will be presented 
in a future work (Z. L. Uhm \& B. Zhang, 2015, in preparation).

%
%

\acknowledgments
This work is supported by China Postdoctoral Science Foundation through 
Grant No. 2013M540813, and National Basic Research Program (``973'' Program) of China 
under Grant No. 2014CB845800.

%
%


%
%

\end{document}